\newlength{\absize}
\documentclass[12pt]{article}
\renewcommand{\baselinestretch}{1.5}

\begin{document}
\thispagestyle{empty}
\pagestyle{empty}
\renewcommand{\thefootnote}{\fnsymbol{footnote}}
\newcommand{\starttext}{\newpage\normalsize
 \pagestyle{plain}
 \setlength{\baselineskip}{3ex}\par
 \setcounter{footnote}{0}
 \renewcommand{\thefootnote}{\arabic{footnote}}
 }
\newcommand{\preprint}[1]{\begin{flushright}
 \setlength{\baselineskip}{3ex}#1\end{flushright}}
\renewcommand{\title}[1]{\begin{center}\LARGE
 #1\end{center}\par}
\renewcommand{\author}[1]{\vspace{2ex}{\Large\begin{center}
 \setlength{\baselineskip}{3ex}#1\par\end{center}}}
\renewcommand{\thanks}[1]{\footnote{#1}}
\renewcommand{\abstract}[1]{\vspace{2ex}\normalsize\begin{center}
 \centerline{\bf Abstract}\par\vspace{2ex}\parbox{\absize}{#1
 \setlength{\baselineskip}{2.5ex}\par}
 \end{center}}

\newcommand{\rep}{representation}
\newcommand{\tr}{\mathop{\rm tr}}
\newcommand{\cO}{{\cal O}}
\newcommand{\cL}{{\cal L}}
\newcommand{\cD}{{\cal D}}
\newcommand{\cA}{{\cal A}}
\newcommand{\cM}{{\cal M}}
\newcommand{\cJ}{{\cal J}}
\newcommand{\cK}{{\cal K}}
\newcommand{\cQ}{{\cal Q}}
\newcommand{\cF}{{\cal F}}
\newcommand{\psib}{{\bar\psi}}
\newcommand{\Psib}{{\bar\Psi}}

\newcommand{\half}{{1\over2}}
\newcommand{\gtrsim}
{\raisebox{.2em}{$\rlap{\raisebox{-.5em}{$\;\sim$}}>\,$}}
\newcommand{\ltsim}
{\raisebox{.2em}{$\rlap{\raisebox{-.5em}{$\;\sim$}}<\,$}}
\newlength{\eqnparsize}
\setlength{\eqnparsize}{.95\textwidth}
\newcommand{\eqnbox}[1]{\parbox{\eqnparsize}{\bf\vskip.25ex
 #1\vskip1ex}}
\newcommand{\PSbox}[3]{\mbox{\rule{0in}{#3}\includegraphics{#1}\hspace{#2
 }}}
\def\spur#1{\mathord{\not\mathrel{#1}}}
\newcommand\etal{{\it et al.}}
\def\arctanh{\mathop{\rm arctanh}\nolimits}
\def\sech{\mathop{\rm sech}\nolimits}
\newcommand{\Ket}[1]{\left| #1 \right\rangle}

\setlength{\jot}{1.5ex}
\newcommand{\figsize}{\small}
\renewcommand{\bar}{\overline}
\font\fiverm=cmr5
\input prepictex
\input pictex
\input postpictex
\input{psfig.sty}
\newdimen\tdim
\tdim=\unitlength
\def\stpltsmbl{\setplotsymbol ({\small .})}
\def\bsmbl{\setplotsymbol ({\Huge .})}
\def\tarrow{\arrow <5\tdim> [.3,.6]}
\def\barrow{\arrow <8\tdim> [.3,.6]}

\setcounter{bottomnumber}{2}
\setcounter{topnumber}{3}
\setcounter{totalnumber}{4}
\renewcommand{\bottomfraction}{1}
\renewcommand{\topfraction}{1}
\renewcommand{\textfraction}{0}

\def\draft{\renewcommand{\label}[1]{{\quad[\sf ##1]}}
\renewcommand{\ref}[1]{{[\sf ##1]}}
\renewenvironment{thebibliography}{\section*{References}}{}
\renewcommand{\cite}[1]{{\sf[##1]}}
\renewcommand{\bibitem}[1]{\par\noindent{\sf[##1]}}}

\def\theequation{\thesection.\arabic{equation}}
\preprint{\#HUTP-01/A013\\ BUHEP-01-4\\
LBNL-47614\\ UCB-PTH-01-09\\}
\title{Anomalies on Orbifolds}
\author{
 Nima Arkani-Hamed,\thanks{\noindent{\tt arkani@bose.harvard.edu}\\
 \indent Permanent address: Department of Physics,
 University of California,
 Berkeley, CA 94720}
 Andrew G. Cohen\thanks{\noindent{\tt cohen@andy.bu.edu}\\
 \indent Permanent address: Physics Department,
 Boston University,
 Boston, MA 02215
 }
 and
 Howard~Georgi,\thanks{\noindent \tt georgi@physics.harvard.edu}
 \\ \medskip
 Lyman Laboratory of Physics \\
 Harvard University \\
 Cambridge, MA 02138
 }
\date{2/10}
\abstract{We discuss the form of the chiral anomaly on an $S^1/Z_2$
 orbifold with chiral boundary conditions. We find 
 that the 4-divergence of the higher-dimensional current
 evaluated at a given point in the extra dimension is proportional to
 the probability of finding the chiral zero mode there.
 Nevertheless the anomaly,
 appropriately defined as the five dimensional divergence of the
 current, lives entirely on the orbifold fixed planes and is
 independent of the shape of the zero mode.
 Therefore long distance four dimensional anomaly cancellation
 ensures the consistency of the higher dimensional orbifold theory.
 } 

\starttext

\setcounter{equation}{0}

\section{Introduction\label{sec:introduction}}

Theories involving spatial dimensions beyond the four of ordinary
experience have long been of interest to physicists. Especially
interesting are higher dimensional theories on
orbifolds~\cite{Dixon,Pomarol}, spaces obtained from compact extra
dimensions by dividing by a discrete symmetry. Such a compact theory
may be viewed at low energies as an effective 4-dimensional theory
involving a spectrum of Kaluza-Klein particles. For an orbifold
compactification, the resulting theory may be chiral and the
corresponding gauge theory anomalous~\cite{ABJ}.  To ensure
consistency of the orbifold gauge theory, the anomalies must
cancel. It is therefore interesting to ask what the conditions for
anomaly cancellation look like from the effective theory point of
view. In particular, does cancellation of the 4-dimensional anomaly in
the effective theory imply conservation of the corresponding
5-dimensional current? This is reasonable in the anomaly's avatar as
an infrared phenomenon.
If not, additional restrictions on the low energy
theory beyond conventional anomaly cancellation might be required.

How could a 5-dimensional anomaly,
a failure of 5-dimensional current conservation, remain after
4-dimensional anomaly 
cancellation? If the orbifold theory has an anomaly throughout the
bulk, satisfying anomaly cancellation conditions in the effective
4-dimensional theory would in general not suffice to ensure bulk
anomaly cancellation. Even if the 5-dimensional anomaly is localized
at the orbifold fixed points, if this anomaly depends on details of
the bulk physics, such as the wave functions of the Kaluza-Klein
modes, 4-dimensional anomaly cancellation would again be insufficient
to ensure bulk anomaly cancellation.

We will demonstrate that while orbifold compactification can introduce
anomalous charge non-conservation, it is localized at the orbifold
fixed points. To see this, we must compute the full 5-dimensional
divergence of the 5-dimensional current. The 4-dimensional divergence
gives a contribution to the anomaly that is non-zero in the bulk and
depends on the shape of the zero modes. However, the 5-divergence has an
additional piece that is the 
extra dimensional derivative of
the extra component of the current, and this piece precisely cancels 
the anomaly in the bulk. The complete 
anomaly is independent of the
details of the bulk physics.
For example, for a 5-dimensional fermion coupled to an
external gauge potential $A_C(x,x_4)$ on an $S^1/Z_2$ orbifold 
with fixed points at $x_4=0$ and $x_4=L$ we will
find\footnote{We use a capital Roman letter like $C$ to refer 
 generically to all dimensions, $C=0, 1, 2, 3$ or $4$, and a Greek
 letter like $\mu$ to refer specifically to the usual four
 dimensions.}
\begin{equation}
 \partial_C J^C(x,x_4) = \frac{1}{2} \Bigl[\delta(x_4) +
 \delta(x_4-L)\Bigr] \cQ\,.
 \label{5anomaly}
\end{equation}
where $J^C$ is the 5-dimensional current and $\cQ$ is just the usual
4-dimensional chiral anomaly for a charged Dirac
fermion in an external gauge potential $A_\mu(x,x_4)$:
\begin{equation}
 \cQ = \frac{1}{16\pi^2} F_{\mu\nu} \tilde F^{\mu\nu}
 \label{anomaly}
\end{equation}
This expression has no dependence on the details of the Kaluza-Klein
mode decomposition. This implies that cancellation of the
4-dimensional anomaly is sufficient to eliminate the 5-dimensional
anomaly. 

\setcounter{equation}{0}
\section{The 4-dimensional anomaly\label{sec:4-anomaly}}

We begin with a brief review of a free Dirac fermion of mass $m$ in
four dimensions 
coupled to an external gauge potential $A^\mu(x)$. The classical
equations of motion for the fermion lead to the na\"\i{}ve equation
for the divergence of the axial current $j^\mu_5 = \psib
\gamma^\mu\gamma^5\psi$:
\begin{equation}
 \partial_\mu j_5^\mu + 2im\,\psib\gamma^5\psi = 0 \,.
\end{equation}

However the corresponding operator equation in the quantum theory
receives a famous correction, the anomaly:
\begin{equation}
 \partial_\mu j_5^\mu + 2im\,\psib\gamma^5\psi = Q
 \,.
 \label{axialanomaly}
\end{equation}
with $Q$ given by (\ref{anomaly}).

It is worthwhile to consider the expectation value of this equation in
the presence of the external gauge potential. Note that a non-zero
expectation value for the divergence of the current would require a
pole at $p^2=0$ in the expectation value of the current itself. For a
massive fermion there is no state to produce such a pole, and
consequently the expectation value of the divergence of the axial
current must vanish. The operator equation (\ref{axialanomaly}) then
implies
\begin{equation}
 2im\,\langle \psib\gamma^5\psi \rangle = Q \,.
\label{massiveanomaly}
\end{equation}

Since this theory involves only an external gauge field we are free to
contemplate the above equations for the case of a single Weyl fermion
rather than a Dirac fermion. In this case the fermion is massless, and
the mass term in (\ref{axialanomaly}) vanishes. Also the anomaly of
the 2-component Weyl fermion is
half that of a 4-component Dirac fermion:
\begin{equation}
 \mbox{(Weyl fermion)} \qquad \partial_\mu j^\mu = \frac{1}{2}Q \,. 
\end{equation}
For the Weyl fermion the expectation value of the divergence of the current
does not vanish---the massless chiral fermion is exactly the state
needed to produce the pole in the expectation value of the axial
current:
\begin{equation}
 \mbox{(Weyl fermion)} \qquad \langle \partial_\mu j^\mu \rangle =
 \frac{1}{2}Q \,. 
\end{equation}

In our subsequent discussion we will encounter a theory with many
fermions with various masses $M_i$ and chiral charges $q_i$,
interacting with gauge potentials with complicated non-diagonal
chiral couplings to the fermions $A^\mu_{ij}$. The generalization of
(\ref{axialanomaly}) is straightforward:
\begin{equation}
 \partial_\mu J_5^\mu + 2i\,\psib M\gamma^5\psi = \half\cQ
 \,,
 \label{axialanomaly2}
\end{equation}
where $\cQ$ is the trace of the chiral charge with the external
gauge fields:
\begin{equation}
 \cQ = \frac{1}{16\pi^2}\, \tr q\, F\cdot\tilde F\,.
\end{equation}
A non-zero expectation value of the divergence of the current still
requires a pole in the current itself. The arguments of
Coleman-Grossman~\cite{Coleman} show that only massless modes can
produce such a pole, and consequently
\begin{equation}
 \langle \partial_\mu J^\mu_5\rangle = \frac{1}{32\pi^2}\, \tr \Bigl(\,P_0\,
 q\,P_0\, F\,P_0\cdot\tilde F\,\Bigr)\,. 
\label{masslessanomaly}
\end{equation}
where $P_0$ is the projector onto the massless sector of the theory.

We will see that the essential difference in the emergence of the
anomaly for massless and massive fermions is reflected in the
structure of the 5-dimensional anomaly. Only zero modes associated
with massless fermions in the effective 4-dimensional theory produce
anomalies in the 4-dimensional divergence. However, massive modes can
still affect the anomaly through terms like (\ref{massiveanomaly}).

\setcounter{equation}{0}
\section{Chiral fermions in five dimensions\label{sec:chiral}}

Our starting point is a chiral orbifold model with a 4-component
fermion field in five dimensions coupled to a classical gauge field.
In order to avoid unnecessary clutter in our equations, we will
consider only the abelian example --- the extension to the non-abelian
case is straightforward. The extra dimension, $x_4$, is in the
interval $[0,L]$ and the mass depends on $x_4$. The action is
\begin{equation}
 S=\int\!\! dx\,\int_0^L \!\!dx_4\, \psib\Bigl(i\spur
 D-i\gamma_4D_4-m(x_4)\Bigr)\psi\,,
 \label{f1}
\end{equation}
where
\begin{equation}
 \spur D = \gamma^\mu D_\mu, \qquad D_C=\partial_C+i\,A_C 
 \label{cd1}
\end{equation}
with $A_C$ a classical gauge potential. This theory has a na\"\i{}vely
conserved current
\begin{equation}
 J^C(x,x_4)=\psib(x,x_4)\,\gamma^C\,\psi(x,x_4)\,.
 \label{current}
\end{equation}
The Dirac matrix $\gamma_4$ is related to what is conventionally called
$\gamma_5$ by a factor of $i$,
\begin{equation}
 \gamma_4=-i\gamma_5\,.
 \label{g4to5}
\end{equation}

The key to the model is the orbifold construction that
restricts the physical region in the extra dimension to the interval
$[0,L]$. 
To implement this we extend the fields to functions on the
doubled interval $x_4\in [0,2L)$, and then impose
\begin{eqnarray}
 &&\psi(x,x_4) =\psi(x,2L+x_4)=\gamma_5\,\psi(x,-x_4) \,,
 \label{f0}\\
 &&A_\mu(x,x_4) =A_\mu(x,2L+x_4)=A_\mu(x,-x_4)\,,%
 \label{fa0}\\
 &&A_4(x,x_4) =A_4(x,2L+x_4)=-A_4(x,-x_4)\,,
 \label{fa50}
\end{eqnarray}
In order for the action to be well defined the mass function must
satisfy
\begin{equation}
 m(x_4)=m(2L+x_4)=-m(-x_4)\ .
 \label{fm1}
\end{equation}
It is through the boundary condition (\ref{f0}) that
chirality enters into the theory. Specifically,
(\ref{f0}-\ref{fa50}) imply that the boundaries of the physical
region, $x_4=0,L$, are fixed-points of the orbifold.
If we decompose $\psi$ into chiral components, $\psi_\pm$, where
\begin{equation}
 \psi=\psi_++\psi_-\,,\quad \gamma_5\psi_\pm=\pm\psi_\pm\,,
 \label{f4}
\end{equation}
then (\ref{f0}-\ref{fa50}) are equivalent to defining all the fields on
a circle, $x_4\in[0,2L)$ with $2L$ identified with $0$, but with 
$\psi_+$, and $A_\mu$ symmetric about the fixed points
$x_4=0,L$, and $\psi_-$ and $A_4$ antisymmetric.

The classical Lagrangian (\ref{f1}) and the orbifold boundary conditions
(\ref{f0}-\ref{fa50}) are invariant under gauge
transformations of the form
\begin{eqnarray}
 \psi(x,x_4)&\rightarrow& e^{i\phi(x,x_4)}\,\psi(x,x_4)\,,
 \label{g1}\\
 A_\mu(x,x_4)&\rightarrow& A_\mu(x,x_4)-\partial_\mu\,\phi(x,x_4)\,,
 \label{g2}\\
 A_4(x,x_4)&\rightarrow& A_4(x,x_4)-\partial_4\,\phi(x,x_4)\,.
 \label{g3}
\end{eqnarray}
provided $\phi$ satisfies
\begin{equation}
 \phi(x,x_4) =\phi(x,2L+x_4)=\phi(x,-x_4)\,.
\label{g4}
\end{equation}

The orbifold boundary conditions (\ref{f0}-\ref{fa50}) give rise
to a massless chiral fermion, and we expect
that this gauge symmetry is then broken by the chiral anomaly. We seek the
precise form of the anomaly. In particular, we are interested in the
``shape'' of the anomaly in the extra dimension. Because we have kept
the mass function $m(x_4)$ arbitrary except for the boundary
conditions, we can get very different shapes for the wave function of
the chiral zero mode in the extra dimension. If the anomaly depends on
the shape of the zero mode, and therefore on the mass function, this
would make it difficult to cancel the anomaly in the
5-dimensional theory. We will see explicitly that this does not happen.

\setcounter{equation}{0}
\section{Calculation of the Anomaly\label{sec:anomaly}}

We begin by choosing a gauge\footnote{The gauge
 transformations (\ref{g1}-\ref{g3}) and the boundary conditions
 (\ref{f0}-\ref{fa50}) guarantee that such a gauge exists.}
 in which $A_4=0$.
In this gauge the KK mode wave functions are independent of the
gauge fields.
To decompose the fermion field into KK modes, we
define the functions $\xi_M^\pm(x_4)$ satisfying
\begin{equation}
 \Bigl[-\partial_4+m(x_4)\Bigr]\xi_M^-(x_4)=M\xi_M^+(x_4)\,,\quad
 \Bigl[\partial_4+m(x_4)\Bigr]\xi_M^+(x_4)=M\xi_M^-(x_4)
 \label{7}
\end{equation}
with $M\geq0$ for $\xi_M^+$ and $M>0$ for $\xi_M^-$. The
$\xi_M^\pm(x_4)$ can be chosen real and respectively symmetric and
antisymmetric about the
fixed points $x_4=0$ and $x_4=L$. The $M$s are the masses of the KK
modes and the $\xi_M^\pm(x_4)$ are their wave functions. They form an
orthogonal basis for the functions on the circle $[0,2L)$
respectively symmetric and antisymmetric about the fixed points. 
Alternatively they form an orthogonal basis for the functions on
$[0,L]$ satisfying respectively Neumann and Dirichlet boundary
conditions at the fixed points.
We normalize them such that
\begin{equation}
 \int_0^L \!\! dx_4\,\xi_M^+(x_4)\,\xi_{M'}^+(x_4) =
 \int_0^L \!\! dx_4\,\xi_M^-(x_4)\,\xi_{M'}^-(x_4) = \delta_{MM'} \,.
 \label{norm1}
\end{equation}

The orbifold boundary conditions ensure that no zero mode appears
in the $\xi^-_M$. 
It is nevertheless convenient to introduce $\xi^-_0\equiv 0$ which will
allow us to treat the plus and minus modes more symmetrically. This
mode clearly does not satisfy (\ref{norm1}), but we will never
encounter a formula which involves the norm of this function.
Now we expand the fermion fields in the $\xi_M$s
\begin{equation}
 \psi_{\pm}(x,x_4)=
 \sum_M\,\psi_{M\pm}(x)\,\xi_M^\pm(x_4)\,,
 \label{f6}
\end{equation}
where 
\begin{equation}
 \psi_{M\pm}(x)=\int_0^Ldx_4\,\xi_M^\pm(x_4)\,\psi_{\pm}(x,x_4)\,.
 \label{f7}
\end{equation}

Inserting (\ref{f6}) into the action, (\ref{f1}), gives
\begin{eqnarray}
 &\displaystyle S=\int\!\! dx\,\int_0^L \!\! dx_4\,
 \Biggl( \sum_M\Biggl[\psib_{M+}(x)\,\xi_M^+(x_4)
 +\psib_{M-}(x)\,\xi_M^-(x_4)\Biggr]
 &
 \\
\nopagebreak[4]
 &\Bigl(i\spur\,\partial -\spur A
 -\gamma_5\partial_4-m(x_4)\Bigr)& \label{kk2}\nonumber
 \\
\nopagebreak[4]
 &\displaystyle
 \sum_{M'}\Biggl[\psi_{M'+}(x)\,\xi_{M'}^+(x_4)
 +\psi_{M'-}(x)\,\xi_{M'}^-(x_4) \Biggr]\Biggr)
 &\nonumber\\
 &\displaystyle =\int\!\! dx\,
 \Biggl(\sum_M\psib_{M}(x)\,\Bigl(i\spur\,\partial-M\Bigr)\,
 \psi_{M}(x)
 &
 \label{kk3}
 \\
\nopagebreak[4]
 &\displaystyle 
 -\sum_{M,M'}\psib_{M'+}(x)\,\spur A_{M'M}^+(x)\,
 \psi_{M+}(x)
 -\sum_{M,M'}\psib_{M'-}(x)\,\spur A_{M'M}^-(x)\,
 \psi_{M-}(x)
 \Biggr)&
 \label{kk4}\nonumber
\end{eqnarray}
where
\begin{equation}
 \psi_M(x)=\psi_{M+}(x) +\psi_{M-}(x)
 \label{dirac}
\end{equation}
is a 4-component 4-dimensional Dirac field for each $M>0$ and 
equal to $\psi_0^+$ for $M=0$, and
\begin{equation}
 A^{\mu\pm}_{M'M}(x)=
 \int_0^L\,dx_4\,\xi_{M'}^\pm(x_4)
 \,\xi_M^\pm(x_4)\,A^\mu(x,x_4)\,.
 \label{matrixA}
\end{equation}
Using this mode decomposition the components of the current $J^C$
become 
\begin{eqnarray}
 &\displaystyle J^\mu(x,x_4) = \sum_{M',M}\,\Biggl(\xi_{M'}^+(x_4)\,\xi_M^+(x_4)
 \,\psib_{M'+}(x)\,\gamma^\mu\, \psi_{M+}(x)
 &
\nonumber \\
\nopagebreak[4]
 &\displaystyle +\xi_{M'}^-(x_4)\,\xi_M^-(x_4)
 \,\psib_{M'-}(x)\,\gamma^\mu\, \psi_{M-}(x)\Biggr)\,,
 &
 \label{kk6}
 \\
 &\displaystyle
 J^4(x,x_4) = \sum_{M', M}\,\Biggl(\xi_{M'}^+(x_4)\,\xi_M^-(x_4)
 \,\psib_{M'+}(x)\,i\gamma_5\, \psi_{M-}(x)
 &
\nonumber \\
\nopagebreak[4]
 &\displaystyle
 +\xi_{M'}^-(x_4)\,\xi_M^+(x_4)
 \,\psib_{M'-}(x)\,i\gamma_5\, \psi_{M+}(x)\Biggr)
\,,
 &
 \label{kk8}
\end{eqnarray}

We can write all this in a useful matrix notation by collecting the
$\psi_M(x)$ into a column vector $\Psi(x)$, the gauge potentials into
matrices $\cA^{\mu\pm}$ with matrix elements $A_{M'M}^{\mu\pm}$ and
introducing the mass
matrix $\cM$ with matrix elements $M\delta_{M'M}$.
Then
\begin{eqnarray}
 &\displaystyle S=\int\!\! dx\,
 \Psib(x)\,\Bigl(i\spur\,\partial-\spur\!\cA-\cM\Bigr)\,\Psi(x)
 &
 \label{mmk3}
\end{eqnarray}
where
\begin{equation}
 \spur\!\cA=\spur\!\cA^+\,P_++\spur\!\cA^-\,P_-
 \label{mma}
\end{equation}
and as usual
\begin{equation}
 P_\pm={1\pm\gamma_5\over2}\,.
 \label{ppm}
\end{equation}
Evidently, this is a theory of a single massless Weyl fermion and an
infinite tower of Dirac fermions, interacting with classical gauge
field $\cA^\mu$ with a complicated matrix coupling in the flavor space
of the 4-dimensional fermion fields. Note that these couplings
are chiral.

A similar matrix notation simplifies the current. Define
\begin{equation}
 \Xi(x_4)=\Xi^+(x_4)\,P_+ +
 \Xi^-(x_4)\,P_-\,,
 \label{mmx}
\end{equation}
with $\Xi^\pm(x_4)$ the matrices with matrix elements
\begin{equation}
 \left[\Xi^\pm(x_4)\right]_{M'M} = \xi_{M'}^\pm(x_4) \xi_M^\pm(x_4)\,,
 \label{mmxp}
\end{equation}
and
\begin{equation}
 \Omega(x_4)=\Omega^+(x_4)\,P_+ +
 \Omega^-(x_4)\,P_-\,,
 \label{mmo}
\end{equation}
with $\Omega^\pm(x_4)$ the matrices with matrix elements
\begin{equation}
 \left[\Omega^\pm(x_4)\right]_{M'M} =
 \xi_{M'}^\mp(x_4)\,\xi_M^\pm(x_4)
 \,.
 \label{mmop}
\end{equation}
In this notation the 5-dimensional current (\ref{current}) is 
\begin{equation}
 J^\mu(x,x_4)= \Psib(x) \,\gamma^\mu\,\Xi(x_4)\, \Psi(x)\,,\quad\quad
 J^4(x,x_4)=\Psib(x) \,i\gamma_5\,\Omega(x_4)\, \Psi(x)\,.
 \label{jm1}
\end{equation}

Clasically, this current is conserved. This is obvious from the
starting point, but we can see it directly in this notation by
separately computing the derivative of $J^4$ and the
4-dimensional divergence of $J^\mu$. 
The derivative of $J^4$ is
\begin{eqnarray}
 \partial_4\,J^4(x,x_4)= 
 \Psib(x) \,i\gamma_5\,\partial_4\Omega(x_4)\, \Psi(x)
 \label{cj2}
\end{eqnarray}
and using the equations for the KK wave functions gives
\begin{eqnarray}
 \displaystyle
 \partial_4\Omega^+(x_4)&=&-\cM\,\Xi^+(x_4)+\Xi^-(x_4)\,\cM\,,
 \label{cj3}
 \\
 \displaystyle 
 \partial_4\Omega^-(x_4)&=&\cM\,\Xi^-(x_4)-\Xi^+(x_4)\,\cM\,.
 \label{cj4}
\end{eqnarray}

The 
4-dimensional divergence is 
\begin{eqnarray}
 &\displaystyle \partial_\mu\,J^\mu(x,x_4)= 
 \Bigl(\partial_\mu\,\Psib(x)\Bigr) \,\gamma^\mu\,\Xi(x_4)\, \Psi(x)
 +\Psib(x) \,\gamma^\mu\,\Xi(x_4)\,
 \Bigl(\partial_\mu\,\Psi(x)\Bigr)
 \label{cj1}
 &\\
\nopagebreak[4]
 &\displaystyle 
 =i\Psib(x)\,\cM \Bigl(\Xi^+(x_4)\,P_++\Xi^-(x_4)\,P_-\Bigr)
 \, \Psi(x)
 -i\Psib(x) \Bigl(\Xi^+(x_4)\,P_-+\Xi^-(x_4)\,P_+\Bigr) \,\cM
 \, \Psi(x)
 &\nonumber
\end{eqnarray}
which exactly cancels the contribution of (\ref{cj2}) and gives
current conservation, at least at the classical level.

The cancellation of the $A^\mu$ dependence in (\ref{cj1}) is not
obvious in this notation, but follows because of the completeness of
the $\xi$ functions. We will need this result below, so we will pause
to discuss it here. Consider, for example, the matrix product
\begin{equation}
 \Xi^-(x_4)\,\cA^{\mu-}(x)\,.
 \label{comp1}
\end{equation}
The important point is that we can write
\begin{equation}
 \Xi^-(x_4)\,\cA^{\mu-}(x)=A^\mu(x,x_4)\, \Xi^-(x_4)\,.
 \label{comp2}
\end{equation}
To see this, look at the $M'$-$M$ matrix element (both $M'$ and $M$
non-zero because we are looking at $\Xi^-$),
\begin{eqnarray}
 &\displaystyle \sum_{M''>0}\,\xi_{M'}^-(x_4)\,\xi_{M''}^-(x_4)\,
 \int_0^L\,dx'_4\,\xi_{M''}^-(x'_4)
 \,\xi_M^-(x'_4)\,A^\mu(x,x'_4)
 &
 \label{comp3}
 \\
\nopagebreak[4]
 &\displaystyle 
 =\xi_{M'}^-(x_4)\, \xi_M^-(x_4)\,A^\mu(x,x_4)
 &
 \label{comp4}
\end{eqnarray}
where (\ref{comp4}) follows because the $\xi_M^-(x_4)$ are complete
on the space of functions that vanish at the boundaries, and the
product $\xi_M^-(x_4)\,A^\mu(x,x_4)$ is such a function. Similar
arguments can be used to show that
\begin{equation}
 \cA^{\mu-}(x)\,\Xi^-(x_4) =A^\mu(x,x_4)\, \Xi^-(x_4)\,.
 \label{comp5}
\end{equation}
and
\begin{equation}
 \Xi^+(x_4)\,\cA^{\mu+}(x)= \cA^{\mu+}(x)\, \Xi^+(x_4) =A^\mu(x,x_4)\,
 \Xi^+(x_4)\,.
 \label{comp6}
\end{equation}
Because of (\ref{comp2}), (\ref{comp5}) and (\ref{comp6}), the $A^\mu$ dependence cancels in the 4-dimensional divergence (\ref{cj1}) and the 
5-dimensional current is
classically conserved.

Quantum mechanically, because (\ref{mmk3}) is just a 4-dimensional gauge theory (with an admittedly peculiar gauge field) we know that the current has an
anomalous divergence that is simply the trace of the chiral charge
in the current with the square of the gauge field strength. Thus
the anomaly is equal to
\begin{equation}
 \frac{1}{32\pi^2}\,\tr\Bigl(
 \,\Xi^+(x_4)\, \cF^+(x)\cdot\tilde\cF^+(x)-
 \Xi^-(x_4)\, \cF^-(x)\cdot\tilde\cF^-(x)
 \,\Bigr)
 \label{an1}
\end{equation}
where $\cF^{\mu\nu\pm}(x) = \partial^\mu \cA^{\nu\pm}(x) - \partial^\nu
\cA^{\mu\pm}(x)$.
But now arguments precisely analogous to those leading to
(\ref{comp2}) imply that we can rewrite (\ref{an1}) as
\begin{equation}
 \frac{1}{32\pi^2} F(x,x_4)\cdot \tilde F(x,x_4)
 \,\tr\Bigl(
 \,\Xi^+(x_4) -
 \Xi^-(x_4) 
 \,\Bigr)
 \label{an2}
\end{equation}

It remains to calculate the trace in (\ref{an2}). Using (\ref{mmxp})
we can write
\begin{equation}
 \tr\Bigl(
 \,\Xi^+(x_4) -
 \Xi^-(x_4) 
 \,\Bigr)
 =\sum_{M\geq0}
 \xi_M^+(x_4)^2
 -\sum_{M>0}
 \xi_M^-(x_4)^2\,.
 \label{an3}
\end{equation}
To evaluate this, consider 
\begin{equation}
 \Delta(x_4,y_4)\equiv
 \sum_{M\geq0}
 \xi_M^+(x_4)\, \xi_M^+(y_4)
 -\sum_{M>0}
 \xi_M^-(x_4)\, \xi_M^-(y_4)
 \label{an4}
\end{equation}
which reduces to (\ref{an3}) if we set $y_4=x_4$. Because the
$\xi_M^-(y_4)$ functions are antisymmetric at $y_4=0$ while the
$\xi_M^+(y_4)$ are symmetric, we can write
\begin{equation}
 \Delta(x_4,-y_4)
 =\sum_{M\geq0}
 \xi_M^+(x_4)\, \xi_M^+(y_4)
 +\sum_{M>0}
 \xi_M^-(x_4)\, \xi_M^-(y_4)\,.
 \label{an5}
\end{equation}
But taken together, the $\xi_M^+$ and $\xi_M^-$ are a complete set
of functions with periodic boundary conditions on the interval
$[0,2L)$, so we can write\footnote{The $\xi$s are however still
 normalized on the half-circle, and are then $\sqrt{\mbox{twice}}$ as
 large as the 
 eigenfunctions properly normalized on the circle.}
\begin{equation}
 \Delta(x_4,-y_4)=2\sum_N\,\delta(x_4-y_4-2NL)\,.
 \label{an6}
\end{equation}
Thus
\begin{equation}
 \Delta(x_4,x_4)=2\sum_N\,\delta(2x_4-2NL)
 =\sum_N\,\delta(x_4-NL)\,.
 \label{an7}
\end{equation}
Note that these delta functions defined on the physical interval
$[0,L]$ satisfy
\begin{eqnarray}
 \int_0^L \delta(x_4) f(x_4) = \half f(0) \\
 \int_0^L \delta(x_4-L) f(x_4) = \half f(L)
\end{eqnarray}
Restricting to the interval $[0,L]$, we have the final
result that the anomaly is
\begin{equation}
 \partial_C J^C =\half\Bigl[\delta(x_4)+ \delta(x_4-L)\Bigr]\cQ\,.
 \label{an8}
\end{equation}
This result is very gratifying. It shows that there is no anomaly in
the 5-dimensional bulk and that the anomaly on the orbifold fixed
points is entirely independent of the shape of the modes. This implies
that the cancellation of the 4-dimensional anomaly is sufficient to
eliminate the 5-dimensional anomaly.

Also note that the anomaly appears ``split'' between the two fixed 
points---if we integrate over the extra dimension we pick up one-half
of the anomaly of a chiral mode from $x_4=0$ and one-half from
$x_4=L$. Again this is independent of the shape of the chiral zero
mode.

Using (\ref{masslessanomaly}), we can compute the matrix element of
$\partial_\mu J^\mu$. Only the chiral zero mode 
contributes, so 
\begin{equation}
 \langle \partial_\mu J^\mu(x,x_4) \rangle = \half
 \cQ_0 \:\xi_0^+(x_4)^2
\label{spread}
\end{equation}
where $\cQ_0$ is just $\cQ$ calculated with gauge potential
$A_{00}^{\mu+}$. Thus the 4-divergence of the current has a matrix
element varying in the bulk as the square of the zero mode wave
function, but this variation is precisely canceled by the rest of the
5-divergence, to produce (\ref{an8}).

\setcounter{equation}{0}
\section{Anomaly Cancellation}

We close this brief note with an example of anomaly
cancellation. Because the anomaly is independent of the bulk physics,
cancellation of anomalies is also straightforward. If we have a
collection of 5-dimensional fermions, all that is required is that the
zero modes form an anomaly-free representation of the low-energy
4-dimensional gauge group. These zero modes may have completely
different wave functions in the extra dimension. Our analysis in
sections \ref{sec:chiral} and \ref{sec:anomaly} shows that the
5-dimensional anomaly is independent of the details and cancels if the
4-dimensional low energy theory is anomaly free. As an extreme
example, consider a theory with 2 fermions: $\Psi$ and $X$ with charge
$+1$ and $-1$ and piece-wise constant mass terms $m_\Psi(x_4) =
-m_X(x_4)=m$ for $0<x_4<L$ and satisfying the boundary condition
(\ref{fm1}). The zero modes $\psi_0,\chi_0$ have charge $+1,-1$ and
therefore the 4-dimensional low energy theory is anomaly free. But for
large $mL$, these zero modes are concentrated at opposite
boundaries. For large $mL>0$, the zero mode $\psi_0$ is concentrated
near $x_4=0$ while the zero mode $\chi_0$ is concentrated near $x_4=L$
(and {\it vice versa} for $m<0$).  In the limit $mL\rightarrow\infty$,
the non-zero modes are arbitrarily heavy and the zero modes live
entirely on the separate fixed points at $x_4=0$ and
$x_4=L$. Nevertheless, our general analysis shows that the
5-dimensional anomalies must cancel for all $m$.

For simplicity we will show how this works for a 4-dimensional gauge
potential $A^\mu(x)$ which is constant in $x_4$, which corresponds to
turning on only the gauge field zero mode. In this case
$A^\mu_{00}(x)=A^\mu(x)$.  Here we will discuss in detail only the
extreme limit, $mL\to\infty$ and show how our result for the form of
the anomaly can be interpreted in terms of familiar 4-dimensional
results.

First consider the contribution of $\Psi$ to the anomaly. As $mL\to\infty$
the square of the properly normalized zero mode wave function goes to
\begin{equation}
  \psi_0^+(x_4)^2=2\,\delta(x_4)
  \label{mltoinfty}
\end{equation} and the 5-dimensional theory has a chiral fermion bound
to the fixed point $x_4=0$~\cite{Howard}. From (\ref{spread}) we see
that the 
4-divergence is exactly what we expect from a single chiral fermion
localized at $x_4=0$:
\begin{equation}
  \langle \partial_\mu J_\psi^\mu(x,x_4) \rangle = \half
  \cQ \: 2\,\delta(x_4)\,.
  \label{psia}
\end{equation}
 
Since the full anomaly (\ref{an8}) is always evenly split between the
two fixed points, the contribution from $\partial_4 J^4$ must then be
\begin{equation}
  \langle \partial_4 J_\psi^4 \rangle = -\half \cQ\, \delta(x_4) +
  \half \cQ\,  \delta(x_4-L)\,.
\label{cs}
\end{equation}
This has a natural effective theory interpretation. Since the fermion
in the bulk is massive, we should integrate it out. This results in a
Chern-Simons term in the bulk effective action~\cite{CallanHarvey},
whose gauge variation resides entirely at the boundaries $x_4=0,L$,
reproducing (\ref{cs}). The full anomaly is the sum of the
4-dimensional divergence and the variation of the Chern-Simons term,
and this sum is evenly split between the two fixed points at $x_4=0$
and $L$.

For $X$, as $mL\to\infty$ 
the square of the properly normalized zero mode wave function goes to
\begin{equation}
  \chi_0^+(x_4)^2=2\,\delta(x_4-L)\,,
  \label{mltoinfty2}
\end{equation}
the
5-dimensional theory has a chiral fermion bound to the
fixed point $x_4=L$, and the 4-divergence of the current
has the form
\begin{equation}
  \langle \partial_\mu J_\chi^\mu(x,x_4) \rangle = -\half
  \cQ \: 2\,\delta(x_4-L)\,.
  \label{chia}
\end{equation}
The full anomaly (\ref{an8}) is always evenly split between the
two fixed points, and thus the contribution from $\partial_4 J^4$ must 
be the same as that in (\ref{cs}),
\begin{equation}
  \langle  \partial_4 J_\chi^4 \rangle = -\half \cQ\, \delta(x_4) +
  \half \cQ\, \delta(x_4-L)\,.
  \label{cs2}
\end{equation}
Again, this comes from the variation of a
Chern-Simons term in the bulk effective action. Adding (\ref{psia}),
(\ref{cs}), (\ref{chia}) and (\ref{cs2}), we see explicitly the
cancellation of the 
5-dimensional anomaly. 

\section*{Acknowledgements}

H.G. is supported in part by the National Science Foundation under
grant number NSF-PHY/98-02709. A.G.C. is supported in part by the
Department of Energy under grant number
\#DE-FG02-91ER-40676. N.A-H. is supported in part by the Department of
Energy. under Contracts DE-AC03-76SF00098, the National Science
Foundation under grant PHY-95-14797, the Alfred P. Sloan foundation,
and the David and Lucille Packard Foundation.

\end{document}